\documentclass[twocolumn,superscriptaddress,showpacs,amsmath,amssymb,aps,pra,floatfix]{revtex4-1}
\usepackage{graphicx}
\usepackage{bm}
\usepackage{amssymb}
\usepackage{amsmath}
\usepackage{amsthm}
\usepackage{amsfonts}
\usepackage{mathrsfs}
\usepackage{hyperref}

\begin{document}

\title{Enhancement of the security of a practical continuous-variable quantum-key-distribution system by manipulating the intensity of the local oscillator}

\author{Xiang-Chun Ma}\affiliation{Department of Physics, National University of Defense Technology, Changsha 410073, People's Republic of China}

\author{Shi-Hai Sun}\affiliation{Department of Physics, National University of Defense Technology, Changsha 410073, People's Republic of China}

\author{Mu-Sheng Jiang}\affiliation{Department of Physics, National University of Defense Technology, Changsha 410073, People's Republic of China}

\author{Ming Gui}\affiliation{Department of Physics, National University of Defense Technology, Changsha 410073, People's Republic of China}

\author{Yan-Li Zhou}\affiliation{Department of Physics, National University of Defense Technology, Changsha 410073, People's Republic of China}

\author{Lin-Mei Liang}\email{nmliang@nudt.edu.cn}
\affiliation{Department of Physics, National University of Defense Technology, Changsha 410073, People's Republic of China}
\affiliation{State Key Laboratory of High Performance Computing, National University of Defense Technology, Changsha 410073, People's Republic of China}

\begin{abstract}
In a practical continuous-variable quantum-key distribution (CVQKD), the fluctuations of the local oscillator (LO) not only make the normalization of Bob's measurement outcomes difficult, but also can change the signal-to-noise ratio (SNR) of an imperfect balanced homodyne detector (BHD), which may lead the security of a practical system of CVQKD to be compromised severely. In this paper, we propose that the LO intensity can be manipulated by the legitimate parties, i.e., being tuned and stabilized to a required constant value, to eliminate the impact of LO fluctuations and defeat Eve's potential attack on the LO. Moreover, we show that the secret key rate can be increased over a noisy channel, especially the channels of metropolitan QKD networks, by tuning the intensity of LO and thus the SNR of a practical BHD to an optimal value, and we find that, counterintuitively, the requirement on BHD (i.e., high detection efficiency and low electronic noise) can also be reduced in this case. To realize this manipulation, we give a schematic setup which thus can be used to enhance the security of a practical CVQKD system.
\end{abstract}

\maketitle

\textit{Introduction}.---
Continuous-variable quantum-key distribution (CVQKD), which exploits two beams [the signal beam carrying the encoding information and the local oscillator (LO) as a phase reference of the signal state used for balanced homodyne detection] to distribute secret keys between two legitimate parties (Alice and Bob), has received great developments in the past decade \cite{Wee12R}. Since the Grosshans-Grangier protocol (GG02 protocol) \cite{Gro02} was first proposed in 2002, various experimental implementations of CVQKD have been carried out by a few groups \cite{Gro03N,Lan05,Lod07A,Qi07,Xua09,She10,Fos09,Jou12,Jou13N}, which makes the CVQKD more promising and practical.

Currently, the security of CVQKD has been demonstrated theoretically \cite{Gar06,Nav06,Lev09L,Ren09L,Lev13L}, but the practical system's security is still the focus of research, such as the analysis of imperfections and the discovery of loopholes of practical systems \cite{Has08,YM11,Jou12A,Sun12,Hua13,Ma13,Ma13A,Jou13A}. In Refs.~\cite{Has08,YM11}, LO was pointed out to be a vulnerability for the practical system, since it is under Eve's control. Furthermore, in Ref.~\cite{Ma13A}, we showed that LO intensity fluctuations open a loophole for Eve and small fluctuations could compromise the security of the practical CVQKD system severely. Therefore, we point out that LO should be monitored accurately and Bob's measurements should be scaled with the instantaneous intensity value of each pulse of LO. However, scaling each pulse's measurement with the instantaneous value makes Bob's measuring complicated. Besides, the analysis in Ref.~\cite{Ma13A} was carried out with a perfect balanced homodyne detector (BHD), however, for a nonideal BHD, Eve still could intercept partial secret keys as analyzed in the rest of this paper if Bob takes the instantaneous normalization.

In this paper, to guarantee the security of the practical system, we propose an alternative countermeasure to solve this problem; i.e., Bob actively stabilizes LO to defeat Eve's LO intensity attack (LOIA, see details in Ref.~\cite{Ma13A}). Furthermore, based on this countermeasure, we point out that the LO could be manipulated not only by Eve to attack the system, but also by Bob in turn to increase the secret key rate over a noisy channel, due to the fact that the variation of LO can change the signal-to-noise ratio (SNR) of a practical BHD \cite{App07}. Our numerical simulation shows that this manipulation of LO can improve the resistance of a practical system to the channel excess noise and may reduce the requirement on BHD (i.e., high detection efficiency and low electronic noise), just by Bob tuning and stabilizing LO intensity to a required optimal constant value. Before describing this countermeasure, we first analyze the effect of LOIA on a practical imperfect BHD of Bob.

\textit{Nonideal BHD attack}.---
In realistic experiments, the nonideal BHD of Bob is characterized by the detection efficiency $\eta$ and the electronic noise $N_{\text{el}}$. Electronic noise is considered as trust noise (the so-called realistic model \cite{Lod07A}) that Eve cannot exploit to acquire information, and its value is calibrated before key distribution based on measuring the variance of BHD when no beams are present. Hence, when the LO intensity fluctuates during the key distribution, this variance after being scaled with the LO instantaneous intensity value (thus called the normalized electronic-noise variance $N_{\text{el}}$) will vary certainly, and thus the SNR of BHD \cite{App07}.

Specifically, for an ideal BHD, the output of its measurement is \cite{Ma13,Ray95}
\begin{equation}\label{eq:X0}
\hat{x}_\theta\approx|\alpha_{\text{LO}}|(\hat{x}\cos\theta+\hat{p}\sin\theta),
\end{equation}
where $\alpha_{\text{LO}}$ is the amplitude of LO, $\theta$ is the relative phase between the signal and LO, and $\hat{x},\hat{p}$ are the quadratures of the signal state. The rotated quadrature $\hat{q}_{\theta}=\hat{x}\cos\theta+\hat{p}\sin\theta$ can be measured with different $\theta$. Equation (\ref{eq:X0}) characterizes a perfect BHD with unity detection efficiency, negligible losses, perfect balancing, and the ideal mode compatibility between the signal and LO. However, for a nonideal BHD, the measured result reads \cite{Had09}
\begin{equation}\label{eq:X00}
\hat{x}_\theta\approx|\sqrt{\eta}\alpha_{\text{LO}}|(\sqrt{\eta}\hat{q}_{\theta}+\sqrt{1-\eta}\hat{x}_N)+\hat{x}_{\text{el}}.
\end{equation}
Here $\eta$ is the detection efficiency, thus $1-\eta$ represents the total optical loss, $\hat{x}_N$ represents the quadrature of the vacuum mode, $\hat{x}_{\text{el}}$ represents the detection noise of BHD and mainly indicates the electronic noise when LO is sufficiently bright (typically $10^9$ photons per pulse).

From Eq.~(\ref{eq:X00}), we can see that when the output of BHD is scaled with $|\sqrt{\eta}\alpha_{\text{LO}}|$ \cite{Footnote} and made statistics with a sufficient number of pulses, the variance of electronic noise can be described as $N_{\text{el}}=\langle\hat{x}^2_{\text{el}}/(\eta|\alpha_{\text{LO}}|^2)\rangle$. If the LO intensity $|\alpha_{\text{LO}}|^2$ is a constant, the electronic noise can be calibrated as a confirmative constant during the calibration procedure. However, when LO fluctuates in time during the key distribution and we scale Bob's measurement with the LO instantaneous intensity value of each pulse (proposed in Ref.~\cite{Ma13A}) instead of the initial calibrated value, the electronic noise will vary, and especially will decrease when $|\alpha_{\text{LO}}|^2$ becomes large. Note that, in this case, we cannot compute the exact value of $N_{\text{el}}$ because, on one hand, we do not know the value of $\hat{x}_{\text{el}}$ in Eq.~(\ref{eq:X00}) during key distribution, and on the other, hand the probability distribution of $|\alpha_{\text{LO}}|^2$ as a random variable is also unknown or unfixed since LO intensity fluctuations are under Eve's control and she can manipulate them arbitrarily.

\begin{figure}[h]
 \includegraphics[width=\columnwidth]{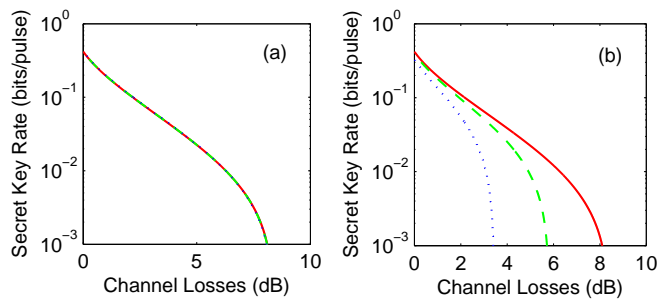}
 \caption{\label{fig:1}(Color online) (a) Secret key rate vs channel losses (dB) with the electronic noise decreasing, i.e., $N_{\text{el}}=0.041$ (solid line), 0.0359 (dashed line), 0.0205 (dotted line), for a fixed channel excess noise $\varepsilon=0.2$ (in shot-noise units). The three curves almost overlapping indicates that the variations of the electronic noise as a trust noise have little impact on the key rate. (b) Truly secret key rate with the case that the decrease of the electronic noise hides the channel excess noise. Solid line indicates $(\varepsilon, N_{\text{el}})$ is (0.2, 0.041), dashed line [$0.2+0.041/(8\eta T)$, 0.0359] and dotted line [$0.2+0.041/(2\eta T)$, 0.0205] where $T$ is the channel transmission, provided that the total noise refereed to the channel input, $\chi_T=\frac{1-\eta T}{\eta T}+\varepsilon+\frac{N_{\text{el}}}{\eta T}$, keeps constant. Bob has no idea about the real electronic noise when LO fluctuates, so if he still considers $\varepsilon=0.2, N_{\text{el}}=0.041$ in the above three cases, he will overestimate his key rates in the later two cases (see text for further details). The curves are plotted for Alice's mode variance $V=40$ (in shot-noise units) with BHD parameters selected as in Ref.~\cite{Lod07A} (i.e., $\eta=0.606, N_{\text{el}}=0.041$).}
\end{figure}

In the realistic model, the electronic noise is considered as trust noise \cite{Lod07A}, and its small variations have little impact on the secret key rate of Alice and Bob as shown in Fig.~\ref{fig:1}(a). However, when the electronic noise varies, especially becomes small, Bob will underestimate the channel excess noise, since he cannot know exactly the electronic noise value when he uses the LO instantaneous intensity value to scale his measurements for each pulse. In addition, this will make the secret key rate compromised as shown in Fig.~\ref{fig:1}(b); i.e., when the electronic noise decreases, Bob will overestimate his secret key rate if he still uses the initial electronic noise to calculate the key rate with the realistic model.

Figure \ref{fig:1}(b) describes the cases that channel excess noise is too high while the real electronic noise decreases drastically, so the secret key rate is compromised severely if the initial calibrated normalized electronic noise is still used in the calculation. Additionally, if Bob classifies his measurements by different values of $|\alpha_{\text{LO}}|^2$ during the parameter estimation procedure, and estimates an average value of normalized electronic noise using the statistic distribution of $|\alpha_{\text{LO}}|^2$, he still cannot estimate the exact secure rates using the realistic model. This is because the mean value of electronic noise is not the real one (cf. the expression of $N_{\text{el}}$) and the statistic probability distribution of $|\alpha_{\text{LO}}|^2$ in the parameter estimation procedure cannot represent the one in the key extraction procedure since it can be arbitrary as mentioned earlier. In this sense, Eve still could fool Alice and Bob by the appropriate manipulation of the intensity of LO.

In practical experiments, however, the fluctuation of LO is small and channel excess noise is also very small. Hence, the electronic noise also varies a bit and has little impact on the key rate. However, in this case, we have to discard pulses with large fluctuation of LO since this large fluctuation will largely change the SNR of a practical BHD. Nevertheless, this will decrease the efficiency of key distribution. Note that the channel excess noise is under Eve's control, so if Eve implements this nonideal BHD attack, LO fluctuations still open a loophole.

\begin{figure}[h]
 \includegraphics[width=.9\columnwidth]{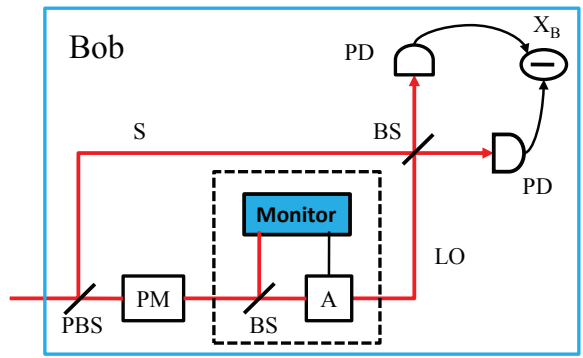}
 \caption{\label{fig:2}(Color online) Schematic setup for stabilizing the intensity of LO. It consists of monitoring each pulse's intensity by splitting a small part of it and using an amplifier or attenuator to amplify or attenuate each pulse which thus stabilizes each pulse's intensity identical with the initial calibrated value. PBS: polarization beam splitter; PM: phase modulator for selecting $\hat{x}$ or $\hat{p}$ quadrature measurement; BS: beam splitter; A: amplifier or attenuator; PD: photodetector. Dashed-line diagram describes the schematic intensity-tuning and stabilizing system.}
\end{figure}

\textit{Countermeasure}.---
However, if we can stabilize the intensity of LO as a constant before the homodyne detection, the above problem will be solved automatically and Eve's attack proposed in Ref.~\cite{Ma13A} will also be avoided efficiently. Figure \ref{fig:2} gives a schematic setup to stabilize the intensity of each pulse of LO, which consists of monitoring LO by splitting a small part with an asymmetric splitter and amplifying or attenuating the LO intensity based on the monitoring value. For the purpose of a relatively simple implementation, for example, we can use a beam splitter, whose transmission is variable, to tune the LO intensity and stabilize it to a required constant value.

Before the homodyne detection, if each pulse's intensity of LO is identical with the calibrated value or fluctuates in an extremely small range, the security of the practical system can be guaranteed and Eve's attack on LO can be avoided. Additionally, with the proposed setup we can tune the LO intensity to an any permitted constant value, thus the SNR (shot-noise to electronic noise ratio) of BHD can be changed to a desired value, which will bring out some merits for CVQKD over a noisy channel.

\textit{Increase secret keys}.---
As shown in the previous section, tuning the intensity of LO changes the SNR of a practical BHD \cite{App07}, which may improve the tolerance of a practical system to channel excess noise. We begin this analysis by first pointing out that adding some noise, classical or quantum, to the reference partner of reconciliation can increase the secret key rate over a noisy channel regardless of the discrete or continuous variable QKD protocol \cite{Ren05A,Ren07L,Mer13,Gar09,Mad12,Wee12A}. Particularly, in Ref.~\cite{Gar09}, Garc{\'{\i}}a-Patr{\'{o}}n and Cerf proposed a new protocol based on squeezed states and heterodyne detection, which is equivalent to the protocol based on squeezed states and homodyne detection but adding some noise on Bob. This additional noise, as a trust noise which is not induced by Eve, makes the protocol more robust against the channel excess noise because it is more detrimental to Eve than to Bob. As pointed out in Ref.~\cite{Gar09}, there is an optimal added noise on the data of the reference partner of reconciliation for a fixed channel excess noise based on squeezed states and homodyne detection; we show that the same is also true of the protocol with coherent states and homodyne detection (see details in Appendix \ref{sec:Noisy}).

\begin{figure}[h]
 \includegraphics[width=\columnwidth]{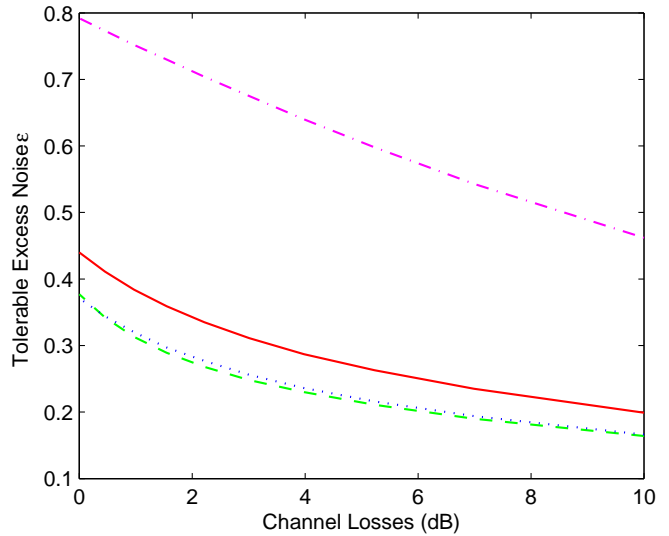}
 \caption{\label{fig:3}(Color online) Tolerable excess noise $\varepsilon$ (in shot-noise units) vs channel losses (in dB) for reverse reconciliation protocol: coherent states and noisy homodyne protocol (solid line), coherent states and perfect homodyne protocol (dashed line), coherent states and heterodyne protocol (dotted line), and squeezed states and noisy homodyne protocol \cite{Gar09} for a comparison (dot-dashed line). The curves are plotted for $V=40$.}
\end{figure}

Hence, in the practical coherent state CVQKD implementation over a noisy channel, we can add some Gaussian noise to Bob's homodyne detection to increase the secret key rate. Over this added noise, the noisy homodyne protocol incredibly improves the resistance to channel excess noise as shown in Fig.~\ref{fig:3}.

\begin{figure}[h]
 \includegraphics[width=\columnwidth]{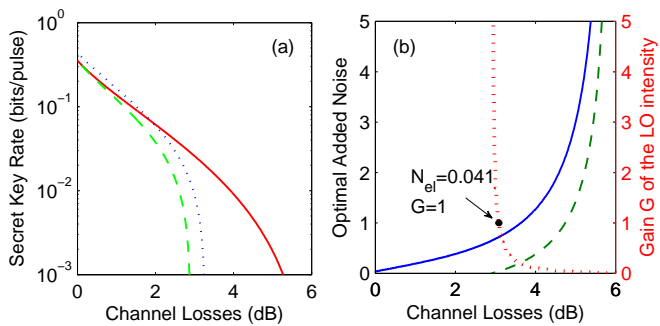}
 \caption{\label{fig:4}(Color online) (a) Secret key rate vs channel losses (dB) with a fixed channel excess noise $\varepsilon=0.25$ for reverse reconciliation protocol, i.e., coherent states and optimal noisy homodyne protocol (solid line), coherent states and perfect homodyne protocol (dashed line), and coherent states and heterodyne protocol (dotted line). (b) Optimal added noise $\chi_D$ on Bob's data vs channel losses (dB) on the left vertical axis (solid line), and dashed line corresponds to the optimal electronic noise. On the right vertical axis, the bold dotted line describes the gain of LO intensity corresponding to the optimal electronic noise (dashed line) according to $N_{\text{el}}=\langle\hat{x}^2_{\text{el}}\rangle/(\eta G|\alpha_{\text{LO}}|^2$) with $\eta=0.606$. The black circle dot pointed at by an arrow indicates that when $N_{\text{el}}=0.041, G=1$.}
\end{figure}

With the optimal added noise $\chi_D$ for a fixed channel excess noise, we plot the secret key rates of different coherent state protocols in Fig.~\ref{fig:4}(a), and find that the noisy homodyne protocol gives a higher secret key rate than the perfect homodyne or heterodyne protocol for higher channel losses or longer transmission distances than a given threshold. The calculation of key rates can be obtained using the standard method \cite{Lod07A,Gar07,Gar09,Wee13} (see Appendix \ref{sec:Calculation} for details).

We also plot the added optimal Gaussian noise [solid line in Fig.~\ref{fig:4}(b)], which can be realized with three methods (see details in Appendix \ref{sec:Noisy}), as a function of channel losses for a fixed channel excess noise $\varepsilon=0.25$. Particularly, we plot the corresponding electronic noise of a practical BHD [dashed line in Fig.~\ref{fig:4}(b)] for a fixed detection efficiency $\eta=0.606$ (selected as Ref.~\cite{Lod07A}) to realize the case of adding optimal Gaussian noise. It is counterintuitive that, for higher channel losses than a given threshold, electronic noise can be beneficial for increasing the secret key rate in CVQKD with the reverse reconciliation (RR) protocol over a noisy channel. As the authors of Ref.~\cite{Gar09} proposed, instead of using a random number generator to generate the noise $\chi_D$, Bob can do it physically by tuning the efficiency of his detector. However, Bob can also realize it by tuning the electronic noise in some limited range, or simultaneously tuning the efficiency and the electronic noise to match the added noise $\chi_D$.

As previous sections analyzed, we can change the SNR or the normalized electronic noise of BHD by tuning the LO intensity. In Fig.~\ref{fig:4}(b), we plot the gain $G$ of LO corresponding to the optimal electronic noise on the right vertical axis (see bold dotted line) and select $G=1$ when the electronic noise is $0.041$ (black circle dot). However, for the higher intensity of LO, the BHD will saturate and thus cannot work, but decreasing less than one order of electronic noise (i.e., $N_{\text{el}}\geq 0.0041, G\leq 10$) may be still realistic in the current setup \cite{Lod07A,Qi07,YM11}. Note that, in this case, we must be very careful since Eve may implement a \textit{saturation attack} \cite{Qin13} on the practical BHD. Hence, we must monitor the mean of Bob's data carefully to avoid this attack (see details in Ref.~\cite{Qin13}). In this sense, the improved performance by tuning LO intensity to higher intensity may be limited in this region. By contrast, for the lower intensity of LO, the noise of BHD may be higher than what Fig.~\ref{fig:4}(b) shows, because there are other excess noises not controlled by Eve, except the electronic noise in the BHD; i.e., the SNR of BHD is not linear dependent on the intensity of LO when LO is not strong and the other excess noise can be still taken as a trust noise. Consequently, with the setup of Fig.~\ref{fig:2}, we can always tune the intensity of LO in the calibration procedure to get the optimal trust added noise supplied by BHD to improve the performance of a practical system over a noisy and lossy channel, and stabilize it on a constant value to avoid Eve's attack on LO during the key distribution. Although this tuning or improvement may be limited, it is still worth testing in practical experiments.

\textit{Discussion and open problem}.---
In the previous sections we have described a schematic setup that can tune the intensity of each pulse of LO to a desired value thus stabilize LO. If we can amplify the intensity of each pulse without changing its phase, the intensity of LO sent by Alice cannot be so strong, which thus could reduce the crosstalk between the signal and LO when they are multiplexed into a same optical fiber. The channel excess noise induced by this crosstalk can be reduced dramatically, which also improves the performance of a practical system of CVQKD. Besides, a general amplifier or attenuator may introduce some phase-insensitive noise, which still can be measured and thus regarded as trust noise to Bob's data. But in this case, we should carefully select these components, avoiding that the introduced extra noise by them exceeds the optimal trust added noise.

Recently, it has been reported that the coherent state CVQKD system was integrated into cryptographic QKD networks and passed some field tests, which confirms the integrability and reliability of it \cite{Fos09,Jou12}. For a field implementation, the channel excess noise is much higher than that of the experimental test and LO also fluctuates dramatically for a long transmission distance. Hence, our proposed stabilization system may be useful and practical for this case, especially for the metropolitan QKD network. Moreover, since the electronic noise may be beneficial for the CVQKD RR protocol over a noisy channel, the requirement on a practical BHD (high efficiency and low electronic noise) can be reduced for high channel excess noise, and in this circumstance the SNR of BHD is not necessarily much higher, which means that the LO intensity can be much lower than the typical value $10^9$ photons per pulse. In other words, we only need to let the SNR (here the noise includes the equivalent noise induced by the detection efficiency) of a practical BHD match the channel excess noise for the RR protocol. It can be realized just by tuning the detection efficiency and LO intensity as the above analyzed for a given BHD, though its performance may be limited for low channel excess noise.

However, we point out that there still exist two open problems to be discussed. First, the above schematic setup needs to be verified experimentally with current facilities and technologies, however, stabilizing each pulse to a constant level may be a big challenge, so seeking good methods will be an interesting and useful target. Second, since the electronic noise may be beneficial for increasing the secret key rate over a noisy channel, it will be very interesting whether the measurement of this BHD is valid or accessible when the normalized electronic noise is close to or even higher than the shot-noise level [as shown in Fig.~\ref{fig:4}(b)], which still needs the experimental verification.

\textit{Conclusion}.---
We analyzed the effect of the LO fluctuation on a practical imperfect BHD in details, and find that LO fluctuation can affect the SNR of BHD, which cannot be measured and is thus unknown to Bob. This could compromise the security of a practical system of CVQKD. Hence, Bob needs to scale his measurements with the instantaneous LO intensity value when LO fluctuates in a extremely small range and select the smallest electronic noise of BHD in this range as the trust noise to calculate the secret key rate. So pulses with a large fluctuation of LO still need to be discarded due to its big change on the SNR of BHD, which thus reduces the efficiency of key distribution. However, we proposed a schematic stabilization system to stabilize the intensity of LO, which could solve this problem and avoid Eve's attack on LO. Then, we studied the impact of Gaussian phase-insensitive noise on the secret key rate for a practical CVQKD system with an RR coherent state protocol, and showed that tuning the LO intensity thus changing the SNR of BHD could increase the secret key rate over a noisy channel. Hence, we can enhance the security of a practical CVQKD system, especially the system integrated into the QKD metropolitan networks, which has high channel excess noise and dramatic LO fluctuation, just by the manipulation of LO, i.e., tuning and stabilizing the intensity of LO.

\begin{acknowledgments}
This work is supported by the National Natural Science Foundation of China, Grants No. 61072071, No. 11304390, and No. 11304391. L.-M.L. is supported by the Program for New Century Excellent Talents. X.-C.M. is supported by the Hunan Provincial Innovation Foundation for Postgraduates. X.-C.M. and M.-S.J. acknowledge support from NUDT under Grant No. kxk130201.
\end{acknowledgments}

\appendix
\section{\label{sec:Noisy}Noisy coherent states protocol}
In this Appendix, we detail the description about the entanglement-based (EB) scheme of a noisy coherent-state protocol of CVQKD for RR, which is analogous to the protocol based on squeezed states and noisy homodyne detection.

\begin{figure}[h]
 \includegraphics[width=\columnwidth]{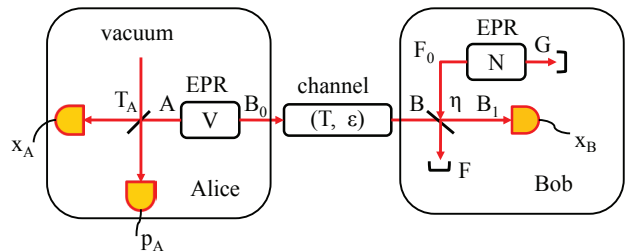}
 \caption{\label{fig:5}(Color online) EB scheme based on coherent states and homodyne detection protocol with added noise on Bob's side for RR over a lossy and noisy channel (transmission $T$ and excess noise $\varepsilon$). Alice makes a heterodyne detection on one half of an EPR state whose variance is V with a 50:50 beam splitter ($T_A=1/2$), and Bob applies a homodyne detection on the other half. Before Bob's homodyne detection, another EPR state of variance $N$, whose one mode is mixed with the signal by a beam splitter (transmittance $\eta$), models the added Gaussian noise.}
\end{figure}

The protocol with coherent states and homodyne detection is shown in Fig.~\ref{fig:5}. Generally, it is used to equivalently describe the practical prepare-and-measure (PM) implementation of CVQKD with a nonideal balanced homodyne detector (BHD) \cite{Lod07A,Qi07}. However, it also can be used to describe the EB scheme of a noisy detection protocol with the coherent states proposed in the main text.

As shown in Ref.~\cite{Gar09}, the added Gaussian phase-insensitive noise can be modeled by the three physical cases: (i) nonideal homodyne detection with efficiency $\eta$ and electronic noise $N_{\text{el}}=(1-\eta)(N-1)$ (see Refs.~\cite{Lod07A,Qi07}); (ii) perfect homodyne detection followed by adding some classical or quantum Gaussian noise of variance $\chi_D=(1-\eta)N/\eta$ \cite{Gar07}; and (iii) any combination of the previous cases giving the same $\chi_D$. Over this added noise, the noisy homodyne protocol incredibly improves the resistance to channel excess noise as shown in Fig.~\ref{fig:3} of the main text.

\section{\label{sec:Calculation}Calculation of the secret key rate}
In this Appendix, we calculate the secret key rate of Gaussian modulation coherent states CVQKD with homodyne detection, heterodyne detection, and noisy homodyne detection protocol, respectively, for RR (see details in Refs.~\cite{Lod07A,Qi07,Gar07,Gar09,Wee13}). For these three cases, the secret key rate can be written by
\begin{equation}
K_{\text{RR}}=\beta I_{AB}-\chi_{BE}.
\end{equation}
Here $I_{AB}$ is the mutual information between Alice and Bob, $\chi_{BE}$ is the Holevo bound between Eve and Bob, and $\beta$ is the reconciliation efficiency. We select $\beta=1$ in the numerical simulation of this paper for simplicity.

For homodyne detection, Alice and Bob's mutual information is given by
\begin{equation}
I_{AB}=\frac{1}{2}\log_2\frac{V_B}{V_{B|A}}=\frac{1}{2}\log_2\frac{V+\chi_C}{1+\chi_C},
\end{equation}
where $V_B=T(V+\chi_C)$ and $V_{B|A}=T(1+\chi_C)$ are Bob's variance and the conditional variance of quadrature measurements, respectively. $\chi_C=(1-T)/T+\varepsilon$ is the total channel added noise referred to the channel input with $T,~\varepsilon$ respective channel transmission and excess noise shown in Fig.~\ref{fig:5}. The Holevo bound $\chi_{BE}$ is given by
\begin{equation}\label{eq:XBE}
\chi_{BE}=S(E)-S(E|B),
\end{equation}
where $S(E)=S(AB)$ can be calculated with the covariance matrix $\gamma_{AB}$ in the EB scheme (see Fig.~\ref{fig:5}), which is given by
\begin{align}
\gamma_{AB}= \left(
              \begin{array}{cc}
                a\mathbf{I} & c\sigma_Z \\
                c\sigma_Z   & b\mathbf{I} \\
              \end{array}
             \right),
\end{align}
where $\mathbf{I}$ and $\sigma_Z$ are the Pauli matrices and $a=V,~b=T(V+\chi_C)$, and $c=\sqrt{T(V^2-1)}$ with $V$ the variance of Alice's mode or the EPR state in Fig.~\ref{fig:5}. Therefore, we have
\begin{equation}\label{eq:SAB}
S(AB)=G\left(\frac{\lambda_1-1}{2}\right)+G\left(\frac{\lambda_2-1}{2}\right).
\end{equation}
Here $G(x)=(x+1)\log_2(x+1)-x\log_2x$, and the symplectic eigenvalues of $\gamma_{AB}$ are given by
\begin{equation}
\lambda_{1,2}=\sqrt{\frac{\Delta\mp\sqrt{\Delta^2-4D^2}}{2}},
\end{equation}
where $\Delta=a^2+b^2-2c^2$ and $D=ab-c^2$. Similarly, $S(E|B)=S(A|B)$ for homodyne detection can be calculated with the covariance matrix $\gamma_A^{x_B}=\text{diag}(a-c^2/b, a)$, where $\text{diag}(,)$ denotes the diagonal matrix with the arguments on the diagonal elements and zeros elsewhere. Hence, $S(A|B)=G[(\lambda_3-1)/2]$ with the symplectic eigenvalue $\lambda_3=\sqrt{a(a-c^2/b)}$.

For heterodyne detection, $I_{AB}$ is given by
\begin{equation}
I_{AB}=\log_2\frac{V_B+1}{V_{B|A}+1}=\log_2\frac{T(V+\chi_C)+1}{T(1+\chi_C)+1},
\end{equation}
The Holevo bound is also given by Eq.~(\ref{eq:XBE}) and $S(E)=S(AB)$ has been obtained by Eq.~(\ref{eq:SAB}). $S(E|B)=S(A|x_B,p_B)$ can be calculated with the covariance matrix $\gamma_A^{x_B,p_B}=\text{diag}[a-c^2/(b+1), a-c^2/(b+1)]$; ie., $S(A|x_B,p_B)= G[(\lambda_4-1)/2]$ with the symplectic eigenvalue $\lambda_4=a-c^2/(b+1)$.

Finally, for noisy homodyne detection, Alice and Bob's mutual information $I_{AB}$ is given by
\begin{equation}
I_{AB}=\frac{1}{2}\log_2\frac{\eta V_B+(1-\eta)N}{\eta V_{B|A}+(1-\eta)N}=\frac{1}{2}\log_2\frac{V+\chi_T}{1+\chi_T},
\end{equation}
where $\chi_T=\chi_C+\chi_D/T$, and $\chi_D=(1-\eta)N/\eta=(1-\eta)/\eta+N_{\text{el}}/\eta$ is the added Gaussian noise. The Holevo bound $\chi_{BE}$ is still given by Eq.(\ref{eq:XBE}) with the unclear term $S(E|B)$, which is given by $S(E|B)=S(AFG|B)= G[(\lambda_5-1)/2]+ G[(\lambda_6-1)/2]$, with the symplectic eigenvalues
\begin{equation}
\lambda_{5,6}=\sqrt{\frac{A\mp\sqrt{A^2-4B}}{2}},
\end{equation}
where $A=\frac{1}{b+\chi_D}[b+aD+\chi_D\Delta]$ and $B=\frac{D}{b+\chi_D}[a+\chi_DD]$ (see details in Refs. \cite{Gar07,Gar09}).
%


\begin{thebibliography}{99}
\bibitem{Footnote} Generally, the measurement outcomes of BHD should be scaled with the amplitude of LO, or alternatively, the variance of measurements is in units with the shot-noise level, because the signal quadratures can be detected thanks to the amplification of LO on them. See details about BHD in Refs.~\cite{Ray95,Had09,Ma13}.
\end{thebibliography}
\end{document}